# Phase field modeling of wetting on structured surfaces

Kaifu Luo[1,*a], Mikko-Pekka Kuittu[1], Chaohui Tong[1], Sami Majaniemi[1,2], and Tapio Ala-Nissila[1,3*b]

[1] Laboratory of Physics, Helsinki University of Technology, P.O. Box 1100, FIN-02015 HUT, Finland

[2] Department of Physics, McGill University, Montreal, Quebec, Canada H3A 2T8

[3] Department of Physics, Box 1843, Brown University, Providence, RI 02912-1843, U.S.A.

ABSTRACT We study the dynamics and equilibrium profile shapes of contact lines for wetting in the case of a spatially inhomogeneous solid wall with stripe defects. Using a phase-field model with conserved dynamics, we first numerically determine the contact line behavior in the case of a stripe defect of varying width. For narrow defects, we find that the maximum distortion of the contact line and the healing length are related to the defect width, while for wide defects, they saturate to constant values. This behavior is in quantitative agreement with experimental data. In addition, we examine the shape of the contact line between two stripe defects as a function of their separation. Using the phase-field model, we also analytically estimate the contact line configuration, and find good qualitative agreement with the numerical results.

a[*] E-mail: luokaifu@yahoo.com
b[*] E-mail: Tapio.Ala-Nissila@tkk.fi

## I. Introduction

Wetting of a solid substrate by liquid plays a very important role in many industrial and natural processes. The dynamics of wetting on clean and smooth surfaces is well understood on the macroscopic scale.[1,2] The equilibrium contact angle at which the meniscus separating a liquid and a gas meets an ideal flat solid surface is related to the various interfacial tensions through the well-known Young's equation, implying a unique contact angle. Experimentally, however, the local contact angles may vary on real substrates. The origin of this effect is that in most situations of practical interests, solid surfaces are rough and chemically heterogeneous. It is well known that chemical defects or roughness of the solid surface may drastically affect the wetting behavior even on a macroscopic level.[1,2] Analysis of wetting properties of liquid on nonideal surfaces has recently become a field with significant technological and industrial applications from oil recovery, printing[3] and biology[4] to micro-fluidics.[5]

One of the most prominent effects of surface inhomogeneities is contact angle hysteresis, where a finite range of static contact angles are observed due to pinning of the contact line in one of the many possible metastable states. Theoretical research on contact angle hysteresis for a liquid on a heterogeneous surface composed of alternatively aligned horizontal stripes confirms the existence of such metastable states.[6] In contrast to the case of an ideal surface, in the presence of heterogeneities the contact line at equilibrium becomes irregular, because it tries to locally deform to find its minimum energy configuration.

A fundamental understanding of these phenomena requires understanding the response of the contact line to isolated inhomogeneities on solid surfaces. Recently, the deformation of the contact line due to a single defect was experimentally investigated in a capillary rise situation. Paterson et al.[7] examined the motion of the contact line past a single circular defect. The results show that there exists a pinning transition. On one hand, weak defects produce a slight distortion of the interface that is released from the defects when the deformation becomes large enough. On the other hand, strong defects produce such a large distortion that the branches of the interface on either side of the defect coalesce to leave an air bubble trapped on the defect. Marsh et al.[8] and Cazabat et al.[9] investigated the contact line configurations on a completely wetted vertical plate near non-wetting vertical stripe defects of varying width. As the width of the stripe defect changes, two regimes are observed: for narrow defects, the behavior of the characteristic healing length correlates with the defect width, while for wide defects, the healing length tends to a constant value. Kanoufi et al.[10] carried out similar experiments, but the saturated healing length was not observed. This may be due to the fact that the width of the stripe was not large enough. In addition, Cubaun et al.[11] examined the shape of the contact line between two isolated defects as a function of their distance, and observed what they called 'individual' and 'collective' pinning regimes.

Theoretically, a variety of approaches on different length scales have been developed to describe the dynamics and equilibrium properties of contact lines. On the microscopic level, molecular dynamics (MD) simulations[12-18] have been used to probe the immediate vicinity of the contact line and to evaluate the validity of the no-slip boundary condition, which is often used in description of simple liquids in contact with a solid surface. In particular, the MD studies[14-18] have shown relative slipping between the fluid and the wall, in violation of the no-slip boundary conditions. The presence of chemical defects and roughness further complicates the dynamics of a contact line. To understand the wetting hysteresis at the molecular scale, Jin et al.[19] have conducted MD simulations of a Wilhelmy plate experiment in which a solid surface is dipped into a liquid bath. The simulation results show that if the surface is microscopically rough, a very irregular local interface shape and an open hysteresis loop corresponding to a history-dependent force emerges. Collet et al.[20] have carried out Monte Carlo simulations to study wetting of a disordered surface. Hysteresis associated with a stick-slip mechanism is measured as a function of disorder and system size. The results show that for a given system size, no measurable hysteresis is found at weak disorder. For a given disorder, even small, a measurable contact angle hysteresis is found above a certain system size.

However, from the experimental point of view perhaps the most interesting results have been obtained using macroscopic approaches[21-26]. Taking into account the elasticity of the contact line, Joanny and de Gennes[25] have derived an equation for the deformation of the contact line resulting from a localized perturbing force, using energy minimization principle. For a stripe defect, their theoretical expression cannot allow for the shape within the defect. Recently, in a different but related approach, Shanahan[26] has derived a more complicated equation, which can allow for the behavior within the defect, too. However, these results are valid only for narrow stripe defects because the maximum distortion of the contact line is always related to defect width. To estimate the contact line configurations resulting from a model of a periodic wettability variation on a vertical solid plate, Schwartz and Garoff[21] have analyzed the energy stored in the meniscus. They have found multiple minima in the energy landscape, deducing that the meniscus motion consists of alternating stick and jump events. Using the same energy, Pomeau and Vannimenus[22] have discussed the contact angle on a heterogeneous surface. Analytic time-dependent calculations of hysteresis have considered in detail the effects of a localized region of different wettability, using various approximations, and with further statistical arguments required for the multiple-defect behavior.

Regarding the dynamical behavior of the contact line, Andersen et al.[27] and Nikolayev et al.[28] have used a macroscopic energy approach to study the dynamics of the pinning of a contact line on a plate with defects. In another recent work, Golestanian et al.[29] have examined the nonequilibrium dynamics of the deformations of a moving contact line on a disordered substrate, taking into account a balance between three different forces: the interfacial force, the frictional force, and a random force caused by the disorder. However, one of the most important physical facts, namely the local conservation of liquid, is at best taken into account only through the lubrication approximation at meniscus level, without obvious manifestation in the contact line equation of motion. It has been shown that the local conservation law is very important for interface dynamics in the closely related imbibition problem.[30-34] In particular, it has been shown in Refs.31-32, that the conservation law leads to a spatially nonlocal equation of motion for the height of the interface, in contrast to the local theories in Refs. 27,28. This completely changes the dynamical behavior of the interface in such cases.

Unfortunately, all of the macroscopic theories discussed above have not explained the experimental results of Marsh et al.[8] and Cazabat et al..[9] Obviously, a complete theoretical description would have to include the contact line shape on a finite-width defect to describe the crossover between narrow and wide defect regimes, and the effects of contact angle hysteresis on the defect may be important as well.

In fact, any interphase boundary is essentially a mesoscopic structure. While the material properties vary smoothly at macroscopic distances along the interface, the gradients in the normal direction are always steep, approaching a molecular scale. This brings about a contradiction between the need for macroscopic description and necessity to take into account microscopic details that influence the motion on larger scales. An important recent advance[30-42] in this respect is the notion that the motion of a contact line can be described by a mesoscopic phase field model.[43] This kind of approach simplifies the interface problem greatly. In many models, handling the boundary conditions for the moving liquid-gas interface causes difficulties. In the phase field model one needs no boundary conditions at all in the liquid-gas interface because interfaces emerge naturally. Moreover, no-slip boundary conditions are not needed in the walls. In this paper, we use the phase-field approach with a model which has been successfully applied to 2D imbibition[30-35]. This model explicitly takes into account local conservation of the liquid phase in the bulk. In the present work, we apply this model to investigate wetting on a wall with stripe defects of varying widths. Both our numerical and analytical results for a single stripe defect indicate that for narrow defects the maximum distortion of the contact line and the healing length are related to the defect width, while for wide defects, they saturate to constant values. These results are in good agreement with the experimental findings of Marsh et al.[8], Cazabat et al.[9] in similar systems. In addition, our numerical results for two stripe defects show with increasing distance between defects there is a crossover from 'collective' pinning of the contact line to individually pinned contact lines at both defects, which has been observed by Cubaud et al.[11]

**II. Phase field model**

The theoretical model used here to study the contact line problem is based on a generalized Landau-type of free energy for the continuum phase field $\phi(\mathbf{r},t)$, which acts as a marker of different phases. The equilibrium values of the field are determined by a free energy functional of the form

$$F(\phi) = \int d\mathbf{r} \left[ \frac{\gamma}{2} |\nabla \phi|^2 + V(\alpha, \phi, A) \right], \qquad (1)$$

where the gradient term describes the energy cost associated due to a spatially varying field. The potential energy term $V(\alpha, \phi, A)$ can be chosen to describe different phases (values of $\phi$) in the system. In the present case, where we want to model a solid wall with a liquid-gas contact line, we choose $V$ to be of the form

$$V = \frac{1}{2}(1+\alpha)\frac{1}{4}(\phi^2 - 1)^2 + \frac{1}{2}(1-\alpha)\frac{\kappa}{2}[\phi - A(\mathbf{r})]^2, \quad (2)$$

which assumes a different form depending on the spatial region under consideration. The fluid and gas phases are characterized by a double well potential ($\alpha = 1$), whereas in the solid phase ($\alpha = -1$), $V$ has only a single well. The solid phase thus has an energy minimum for $\phi = A(\mathbf{r})$, and the liquid and gas phases are described by $\phi = +1$, and -1, respectively. We note that one can also relate the parameters of the phase-field model to the more microscopic density functional theory.[44] In this work, however, our aim is not to model any specific substances in particular, and we choose the parameters in a simple fashion to control the wetting properties of the solid wall, as will be described below.

The liquid-gas contact line on the wall arises naturally through appropriate boundary conditions for the phase field. As usual, the liquid-gas interface is defined by the condition $\phi = 0$. A schematic picture of the geometry of the system is shown in Figures 1 a, b and c. The liquid resides in a 3D reservoir with spatial dimensions $L_x$, $L_y$, $L_z$ in the $x$, $y$, and $z$ directions, respectively. The reservoir is in contact with two vertical walls on the $yz$ plane at $x=0$ and $L_x+1$. A stripe-like defect on the wall at $L_x+1$ is described by a step-like potential $A(x=L_x+1, y, z)$, which gives rise to spatially dependent wetting properties of the wall. A larger value of $A$ indicates a region with preferred wetting, and the limiting values of $A(x=L_x+1, y, z)=+1$ and $-1$ correspond to complete and non-wetting regions, respectively. This can be understood through Young's equation comparing the mesoscopic surface tensions calculated in Appendix B as a function of the 'wall potential' $A$.

The time evolution of the phase-field is given by the continuity equation

$$\frac{\partial \phi(\mathbf{r},t)}{\partial t} + \nabla \cdot \mathbf{j}(\mathbf{r},t) = 0, \quad (3)$$

which guarantees the local conservation of the field. Previous work has shown that this is crucial for a correct description of dynamics in liquids.[30-35] The local differences in the chemical potential $\mu(\mathbf{r},t)$ create a current density $\mathbf{j}(\mathbf{r},t) = -\nabla \mu(\mathbf{r},t)$. The chemical potential can be obtained through the free energy of the system $F(\phi)$ as $\mu = \delta F / \delta \phi$. Inserting Eq. (1) into Eq. (3), we obtain the equation of motion as

$$\frac{\partial \phi(\mathbf{r},t)}{\partial t} = \nabla^2 \left( -\gamma \nabla^2 \phi + \frac{1+\alpha}{2}(\phi^3 - \phi) + \frac{1-\alpha}{2}\kappa(\phi - A(\mathbf{r})) \right). \quad (4)$$

## III. Numerical results

First, we solve Eq. (4) numerically for a system with two solid walls immersed in a fluid phase. One of the walls contains a single rectangular defect of width $L$ extending from the bottom of the wall to the top. The vertical direction is defined as $z$, and the other two directions parallel and perpendicular to walls are $y$ and $x$, respectively, as shown in Figures 1 b and c. We use periodic boundary conditions in the $y$ direction. The initial phase field is set as its equilibrium value in each domain, i.e. the liquid phase has $\phi=1$ in $1\leq x \leq L_x$, $y$, $z=z_0$, the gas phase $\phi=-1$ in $1\leq x \leq L_x$, $y$, $z_0+1\leq z \leq L_z$, the solid phase has $\phi=A(\mathbf{r})=A(y,z)=A$ in $-s+1\leq x \leq 0$, $y$, $0\leq z \leq L_z$ and $L_x+1\leq x \leq L_x+s$, $|y|>L/2$, $0\leq z \leq L_z$ and the defect has $\phi=A(y,z)=B$ in $L_x+1\leq x \leq L_x+s$, $|y|\leq L/2$, $0\leq z \leq L_z$, where $s$ is the width of the wall.

At the solid and gas boundaries, the condition $\nabla\mu=0=-\mathbf{j}$ ensures that there is no mass flow out of the system. At the liquid boundary, two kinds of boundary conditions can be considered. First, in the so-called capillary-driven wetting case $\mu(z=0)=0=const.$, which reflects the presence of an infinite liquid reservoir. Second, in the case of relaxation of the contact line to its equilibrium configuration $\nabla\mu=0$.[30-31] In our simulations, $z_0$ is set to 0 and 20 for the cases of capillary-driven wetting and relaxation of the contact line, respectively.

The Eq.(4) is solved on $L_x \times L_y \times L_z$ three dimensional lattices, with mesh size $\Delta x = \Delta y = \Delta z = 1$. The simple Euler-Scheme is used. Accurate numerical integration requires that the equilibrium liquid-gas interface width must be chosen carefully to be larger than the mesh size. In Eq. (4), the parameter $\gamma$ is related to the correlation length of the liquid-gas interface and gives a measure of its width. Here we have set $\gamma=2$ so that the liquid-gas interface width is $\sqrt{2}$, larger than the mesh size. For the other parameters in the phase-field model, we have used the following values: $\kappa=2$, thickness of the wall $s=5$ and time step $\delta t=0.005$. The system size is chosen as $L_x \times L_y \times L_z = 64 \times 128 \times 64$ unless otherwise stated. We have checked with larger system sizes that our results are not affected by finite-size effects here.

### III.1 Capillary-driven wetting case

### III.1.1 Homogeneous Wall

We first examine the dynamical behavior of a liquid-gas interface starting at rest at the bottom of the system ($z=0$ plane) and moving between two homogeneous solid walls, where $A(x=0, y, z)=A(x=L_x+1, y, z)=A=const.$. In this case, the contact line wets the wall and its profile is a completely straight line. In Figure 2 we show the height of the contact line $H(y, t)$ as a function of time for different values of the wetting parameter $A$. As expected, the height is larger with increasing $A$. In addition, we find that for different values of $A$, $H(y, t)$ shows the eventual crossover towards Washburn equation[45] $H(y, t) \sim t^{0.5}$ for late times.

### III.1.2 One Stripe Defect

Next we consider the case where one of the walls at $x=L_x+1$ contains a stripe defect with $A>B$. In such a case, a typical profile of the meniscus and the liquid-gas contact line in the final equilibrium state for $t \geq 2000$ are shown in Figures 1a and d, respectively. The total contact line distortion $\xi_z = \xi_u + \xi_l$ and the healing length of the distortion $\xi_y$ characterize the effect of the defect on the shape of the contact line.

We first discuss a typical case of capillary-driven wetting where we have chosen $A=0.5$ and $B=-0.5$ such that the defect tends to remain dry while the liquid is pushed up in the regions outside of the defect. In Figure 3a we show the evolution of the contact line for $L=10$. Already for this defect width, the contact line remains almost completely pinned in the middle of the defect. A set of final contact line configurations in equilibrium state with different defect widths are shown in Figure 3b. One of the main features of the contact line configurations is the appearance of two regimes for narrow and wide defects, respectively. This can be seen quantitatively in Figure 3c where we show the distortion lengths $\xi_z, \xi_l$ and $\xi_u$, and the healing length $\xi_y$ as a function of the defect width $L$. All these distortion lengths and the healing length tend to saturate when the defect width becomes larger than $L \approx 14$ and $L \approx 6$, respectively, which occurs due to the relaxation of the contact line on the defect. Namely, only when the defect is wide enough such that $L > max\{\xi_l^*, \xi_u^*, \xi_z^*, \xi_y^*\} \approx 9$, where the asterisks denote the asymptotic values for $L \to \infty$, can the contact line completely relax and the distortion lengths become (almost) independent of $L$. In particular, the healing length $\xi_y$ saturates very rapidly to a constant value of $\xi_y^* = 2.8 \approx \xi_z^*/3$. These results are in excellent agreement with experiments with a non-wetting defect on a completely wet wall,[7,8] where it was found that there are indeed two regimes as the defect width is varied, with $\xi_y^* \approx \xi_z^*/3$.

In Figure 3d, we show a cross-section of the equilibrium meniscus surface in the *xz* plane as measured in the middle of the defect at *y*=0. At *x*=0 where there is no defect on the wall, $\theta_A < \pi/2$ denotes the contact angle between the liquid and the wall, which is completely determined by the wetting property of the wall. At *x*=$L_x$+1, the quantity $\theta_B(L)$ denotes the contact angle between the liquid and the stripe defect.

As shown in Figure 4, the size dependence of the contact angle is more pronounced than for the healing lengths, and only approaches its asymptotic value of 116.5 degrees (as determined by *B*) for the largest stripes studied here.

We have also studied other combinations of the wetting parameters *A* and *B*, and find qualitatively similar results. Final equilibrium configurations for *A*=0.5, *B*=0 and *A*=0, *B*=-0.5 are shown in Figures 5a and b, and the corresponding meniscus profiles in the middle of the defect in Figures 5c and d. In both cases the two regimes for narrow and wide defects can be seen in the distortion lengths. For *A*=0.5, *B*=0, we find that $\xi_y^* = 2.8$ exactly as for case where *A*=0.5, *B*=-0.5. However, now $\xi_z^* = 6.5$, which means that the simple relation $\xi_y^* \approx \xi_z^*/3$ does not hold any more. For the third case where *A*=0, *B*=-0.5, $\xi_y^* \approx \xi_z^* = 2.4$, which indicates that even for the same value of *B*, $\xi_y^*$ is different for different values of *A*.

The sensitivity of the contact line profiles to the wetting properties of the wall and defect is seen in the values of the contact angles in Figures 5c and d. In particular, for *A*=0.5, *B*=0 the angle $\theta_B(L)$ asymptotically approaches $\pi/2$ corresponding to the *B*=0 at the defect, as shown in Figure 4. For the case where *A*=0, *B*=-0.5, on the other hand, $\theta_A \approx \pi/2$, while $\theta_B(L)$ approaches its asymptotic value of 116.5 degrees at the defect boundary, as shown in Figure 4.

**III**.*1.3 Two Stripe defects*

Next, we consider the case where there are two stripe defects of the same width *L*=20 on the wall, and investigate the evolution of the contact line as a function of the distance *d* between the stripes. To compare with the single stripe case, we have chosen *A*=0, *B*=-0.5. In this limit where $d \gg L$, we expect the contact line to accommodate itself to a configuration, which for each stripe is identical to the single, isolated stripe case. Thus, here we concentrate in the limit where $d \leq L$, where the single stripe approximation is not valid. In Figure 6, we show a series of configurations for the evolution of the contact line for different values of *d* ranging from *d*=4 to *d*=*L*=20. From these figures it can be indeed observed that with increasing *d* there is a crossover from 'collective' pinning of the contact line to individually pinned contact lines at both defects. Recently, Cubaud et al.[11] carried an experimental investigation on advancing contact lines of large drops on chemically patterned surface. As a function of the distance between defects, 'individual' pining and 'collective' pinning occur. Our numerical results are in qualitative agreement with their experimental findings.

**III.2 Relaxation of the contact line to its equilibrium configuration**

Finally, we will briefly discuss relaxation of contact line to its equilibrium configuration, where the boundary condition at the reservoir corresponds to a zero gradient of the chemical potential, and in our case we set the initial height of liquid to $z_0$=20. In this case, the profile will simply relax to its equilibrium shape in contrast of the case of a finite flux where it will propagate with a constant velocity.[34,35] The system size is chosen to be $L_x \times L_y \times L_z = 64 \times 200 \times 64$. We focus on case of $A$=0, $B$=-0.5 to facilitate comparison with the capillary-driven wetting case above. In Figure 7a we show the temporal evolution of the contact line profiles for $L$=20. The center of the line $H(y=0, t)$ decreases with time. In Figure 7b we show the final equilibrium contact line profiles for different defect widths, corresponding to Figure 5b for the capillary-driven wetting case. The behavior is rather different from the capillary-driven wetting case, in that the contact line gets fully pinned at $y$=0 only for the largest defects studied here. Second, as can be seen from Figures 7a and c, there is an 'overshoot' of the profiles at the edges of the defect. The magnitude of this effect is dependent on which level-set is chosen to determine the position of the liquid-gas boundary. Here it is defined via $\phi = 0$.

**IV. Analytical results**

One of the advantages in using the phase-field model approach is that it is also possible to estimate the profile of the 1D contact line analytically. Following the approach of Refs. 21, 22 and 28, we can write the height of the contact line (see Appendix A) as

$$h(y) = \int_{-\infty}^{\infty} dy' \int_0^{\infty} dq \frac{\Gamma(h(y), y')\cos[(y-y')t]}{K(q)}, \qquad (5)$$

where $\Gamma(h(y), y')$ and $K(q)$ are spreading and restoring forces, respectively.

It should be pointed out that everything in Eq. (5) is on a macroscopic scale, where the length scale is measured in units of the capillary length. At the wall the value of $\Gamma(y, z) = [\sigma_{sl}(y, z) - \sigma_{sg}(y, z)]/\sigma_{lg}$ is a function of spatial coordinates. Since the defect is symmetric in the $z$ direction we approximate $\Gamma$ to be a function $y$ only. For the key quantity $\Gamma(y)$ in Eq. (5), we can use the phase-field model to estimate the corresponding mesoscopic surface tension terms, as shown in Appendix B. Using the simulated values of $\kappa = 2$ and $\gamma$=2, we get for the phase field model

$$\Gamma(y) = \sqrt{2}\left(\sqrt{2+2\phi(y)} - \sqrt{2-2\phi(y)} - 2\phi(y)\right)\left(-2 + \sqrt{2-2\phi(y)} + \sqrt{2+2\phi(y)}\right)$$
$$+ \frac{1}{3}\left(\left(-1+\sqrt{2+2\phi(y)}\right)^3 + \left(1-\sqrt{2-2\phi(y)}\right)^3\right) \tag{6}$$
$$- \left(\sqrt{2+2\phi(y)} - \sqrt{2-2\phi(y)}\right),$$

where

$$\phi(y) = \begin{cases} \phi_1(y), & \text{for } y < -\frac{L}{2} \\ \phi_2(y), & \text{for } |y| \leq \frac{L}{2} \\ \phi_3(y), & \text{for } y > \frac{L}{2} \end{cases}.$$

Note that $\Gamma$ is dimensionless and can be used directly in the analytic estimate of Eq. (5) although the energy units of our model are different. We approximate the phase field at the wall by replacing it with the phase field profile far inside the wall where the effects of solid-liquid boundary can be neglected.

This profile can be estimated in three different ways. The first way, which we call method 1, is by using the step profile

$$\phi(y) = \begin{cases} \phi_1(y) = A, & \text{for } y < -\frac{L}{2} \\ \phi_2(y) = B, & \text{for } |y| \leq \frac{L}{2} \\ \phi_3(y) = A, & \text{for } y > \frac{L}{2} \end{cases}. \tag{7}$$

A more refined way, called method 2 here, is by requiring that the phase field

$$\phi(y) = \begin{cases} \phi_1(y) = A + a_1 \exp\left(\sqrt{\kappa/\gamma}\, y\right), & \text{for } y < -\frac{L}{2} \\ \phi_2(y) = B + 2b \cosh\left(\sqrt{\kappa/\gamma}\, y\right), & \text{for } |y| \leq \frac{L}{2}, \\ \phi_3(y) = A + a_2 \exp\left(-\sqrt{\kappa/\gamma}\, y\right), & \text{for } y > \frac{L}{2} \end{cases}$$

and its first derivatives $d\phi/dy$ are continuous at the defect edges $y = \pm\frac{L}{2}$. Because of symmetry $a_1 = a_2 = a$, and thus this method gives

$$\phi(y) = \begin{cases} \phi_1(y) = A + \dfrac{(B-A)\exp\left[\sqrt{\dfrac{\kappa}{\gamma}}\left(\dfrac{L}{2}+y\right)\right]}{1+\tanh\left(\sqrt{\dfrac{\kappa}{\gamma}}\dfrac{L}{2}\right)}, & \text{for } y < -\dfrac{L}{2} \\[2ex] \phi_2(y) = B + (A-B)\tanh\left(\sqrt{\dfrac{\kappa}{\gamma}}\dfrac{L}{2}\right)\exp\left(-\sqrt{\dfrac{\kappa}{\gamma}}\dfrac{L}{2}\right)\cosh\left(\sqrt{\dfrac{\kappa}{\gamma}}y\right), & \text{for } |y| \le \dfrac{L}{2} \\[2ex] \phi_3(y) = A + \dfrac{(B-A)\exp\left[\sqrt{\dfrac{\kappa}{\gamma}}\left(\dfrac{L}{2}-y\right)\right]}{1+\tanh\left(\sqrt{\dfrac{\kappa}{\gamma}}\dfrac{L}{2}\right)}, & \text{for } y > \dfrac{L}{2} \end{cases} \quad (8)$$

Another refined way, called method 3 here, is by requiring that the phase field is continuous and the free energy is minimized, which corresponds $d\sigma_{AB}/db = 0$, where $\sigma_{AB}$ is the surface tension of a stripe defect with wall potential $B$. Using Eq. (32) (see Appendix B) we get

$$\sigma_{AB} = 2\sqrt{\dfrac{\gamma}{2}}\int_{\phi_B}^{A}|\tilde{f}(\phi)|d\phi = \sqrt{\kappa\gamma}\left[\phi_0^2 + \dfrac{A^2}{2} - \dfrac{\phi_B^2}{2} + B\phi_B - (A+B)\phi_0\right], \quad (9)$$

where $\phi_B$ is the value of the phase field at $y=0$ and $\phi_0$ is the value of the phase field at $y=L/2$. Minimizing this with respect to parameter $b$ gives

$$\phi(y) = \begin{cases} \phi_1(y) = A + \dfrac{(B-A)\left[\cosh^2\left(\sqrt{\dfrac{\kappa}{\gamma}}\dfrac{L}{2}\right)-1\right]}{2\cosh^2\left(\sqrt{\dfrac{\kappa}{\gamma}}\dfrac{L}{2}\right)-1}\exp\left[\sqrt{\dfrac{\kappa}{\gamma}}\left(\dfrac{L}{2}+y\right)\right], & \text{for } y < -\dfrac{L}{2} \\[3ex] \phi_2(y) = B + \dfrac{(A-B)\cosh\left(\sqrt{\dfrac{\kappa}{\gamma}}\dfrac{L}{2}\right)}{2\cosh^2\left(\sqrt{\dfrac{\kappa}{\gamma}}\dfrac{L}{2}\right)-1}\cosh\left(\sqrt{\dfrac{\kappa}{\gamma}}y\right), & \text{for } |y| \le \dfrac{L}{2} \\[3ex] \phi_3(y) = A + \dfrac{(B-A)\left[\cosh^2\left(\sqrt{\dfrac{\kappa}{\gamma}}\dfrac{L}{2}\right)-1\right]}{2\cosh^2\left(\sqrt{\dfrac{\kappa}{\gamma}}\dfrac{L}{2}\right)-1}\exp\left[\sqrt{\dfrac{\kappa}{\gamma}}\left(\dfrac{L}{2}-y\right)\right], & \text{for } y > \dfrac{L}{2} \end{cases} \quad (10)$$

As to the restoring force, we use $K(q) = \pi\left(1+q^2\right)^{\frac{1}{2}}$, which contains gravity.[22] Here we should point out that $h(y)$ in Eq. (5) is mainly determined by the spreading force $\Gamma(y)$. Therefore the gravity dependent restoring force is a sufficient approximation.

Typical configurations for the contact line profiles using the results from the three methods [see Eqs. (7), (8) and (10)], are shown in Figure 8. The parameters used here are $A=0$ and $B=-0.5$, corresponding to the numerical results in Figure 5b. We find that the profiles are essentially identical as determined from Methods 2 and 3, so only the result for the latter is shown. As can be seen in Figure 8, the analytic estimates are in good agreement with the numerical results. There are again two regimes as the defect width $L$ is varied. For narrow defects, $\xi_z$ correlates with the width of the defects, while for wide defects, it saturates to a constant value $\xi_z^*$. The values obtained for $\xi_z^*$ for the three methods are equal. In terms of the contact line profiles, however, there are some differences. Compared with Methods 2 and 3, as can be seen Figure 8c, Method 1 gives quantitatively different results. As different results are obtained for three methods, we cannot make a detailed comparison of the analytical profiles with the numerically obtained ones.

**IV. Summary and conclusion**

In this work, we have considered the static and dynamical properties of liquid-gas contact lines on solid walls with spatially varying wetting properties. In particular, we have used a modification of a phase-field model, which correctly takes into account the local conservation of the liquid to describe wetting on walls with stripe-like defects. Direct numerical solutions of the phase-field model generate contact line profiles, which are in good agreement with experiments. In particular, for capillary-driven wetting we find that there are two regimes corresponding to narrow and wide defects, where in the latter case the healing length saturates to a constant value in excellent agreement with experiments. We have also considered the shape of the contact line between two stripe defects as a function of their separation. For the case of relaxation of the contact line to its equilibrium configuration, we find that the contact line profiles are qualitatively different from the capillary-driven wetting case, in that they are less strongly pinned and their detailed shapes are different. Finally, we have used a combination of macroscopic arguments and results from the phase-field model to analytically estimate the contact line profiles. Although there is no unique solution, with three different approximate methods we find good qualitative agreement with the numerical solutions.

**Acknowledgement** We thank Martin Rost for his help for numerically convenient representation of the free energy and the surface tensions, and Martin Dube for participating in the early stages of this work. This work has been supported in part by a Center of Excellence grant from the Academy of Finland.

## Appendix

### A. The height of the contact line

The energy stored in the meniscus can be divided into three parts: the liquid-gas interface part ($E_1$), the gravity part ($E_2$) and the wall-fluid part ($E_3$). The last part depends mainly on the shape of the contact line whereas first two depend on also on the shape of the liquid-gas interface. We proceed as follows. Let $Z(x,y)$ be the height of the meniscus and $H(y)=Z(x=0,y)$ the contact line. The first two parts of the energy stored in the meniscus are then given by

$$E_1 = \int_{-\infty}^{\infty} dy \int_0^{\infty} dx \, \sigma_{lg} \sqrt{1+Z_x^2+Z_y^2} - \sigma_{lg}, \tag{11}$$

where $Z_x = \partial Z/\partial x$, $Z_y = \partial Z/\partial y$ and

$$E_2 = \int_{-\infty}^{\infty} dy \int_0^{\infty} dx \, \frac{\rho g Z^2}{2}, \tag{12}$$

By normalizing the length scale by the capillary length $\lambda = \sqrt{\sigma_{lg}/\rho g}$ and the free energy scaled by $\lambda^2 \sigma_{lg}$, we obtain

$$E_1 = \int_{-\infty}^{\infty} dy \int_0^{\infty} dx \sqrt{1+Z_x^2+Z_y^2} - 1, \tag{13}$$

and

$$E_2 = \int_{-\infty}^{\infty} dy \int_0^{\infty} dx \, \frac{Z^2}{2}. \tag{14}$$

The Euler-Lagrange condition $\delta(E_1+E_2)/\delta Z = 0$ then gives

$$Z(x,y) = \frac{\partial}{\partial x} \frac{Z_x}{\sqrt{1+Z_x^2+Z_y^2}} + \frac{\partial}{\partial y} \frac{Z_y}{\sqrt{1+Z_x^2+Z_y^2}}, \tag{15}$$

for the minimum energy configuration. In the case of weak defects, one can assume $H(y)=H_0+h(y)$ with $|h|\ll 1$ and $\partial h/\partial y \ll 1$. Linearizing the meniscus equation by using $Z(x,y)=Z_0(x)+z(x,y)$ with the boundary conditions $Z(x=0,y)=H(y)$, $Z(x=\infty,y)=Z_0(x=\infty)=z(x=\infty,y)=0$, $z(x=0,y)=h(y)$ and $Z_0(x=0)=H_0$ gives

$$z(x,y)\left(1+Z_{0,x}^2\right)^{\frac{5}{2}} = \left(1+Z_{0,x}^2\right)\left[z_{xx}+z_{yy}\left(1+Z_{0,x}^2\right)\right] - 3z_x Z_{0,x} Z_{0,xx}, \tag{16}$$

where the subscript $xx$ and $yy$ denote the corresponding second partial derivatives. Assuming that $Z_0(x)=0$, which corresponds to a contact angle $\pi/2$, the above

equation simplifies to

$$z(x,y) = z_{xx} + z_{yy} \tag{17}$$

In Fourier space this can be written as

$$\tilde{z}_{xx}(x,q) = (1+q^2)\tilde{z}(x,q), \tag{18}$$

where $\tilde{z}(x,q)$ is the Fourier transformation of $z(x,y)$

$$\tilde{z}(x,q) = \frac{1}{\sqrt{2\pi}} \int_{-\infty}^{\infty} dy e^{iqy} z(x,y). \tag{19}$$

Eq. (21) has the solution

$$\tilde{z}(x,q) = \tilde{h}(q)\exp\left(-\sqrt{1+q^2}\,x\right) \tag{20}$$

where

$$\tilde{h}(q) = \tilde{z}(0,q) = \frac{1}{\sqrt{2\pi}} \int_{-\infty}^{\infty} dy e^{iqy} h(y). \tag{21}$$

Eq. (17) above is equal to $\delta e = 0$, with

$$e\{z(x,y)\} = \frac{1}{2}\int_{-\infty}^{\infty} dy \int_{0}^{\infty} dx\left(z^2 + z_x^2 + z_y^2\right) \tag{22}$$

Integrating this by part and using Eq. (17) and the boundary conditions $z(x, y=\pm\infty) = z(x=\infty, y) = z_y(x, y=\pm\infty) = z_x(x=\infty, y)$ we get

$$e\{z\} = -\frac{1}{2}\int_{-\infty}^{\infty} dy z z_x\big|_{x=0} \tag{23}$$

Using the solution in Eq. (23) this becomes in Fourier space

$$e\{h\} = \frac{1}{2}\int_{-\infty}^{\infty} dq \left|\tilde{h}(q)\right|^2 \sqrt{1+q^2}. \tag{24}$$

The wall-fluid part of the energy due to defect can be written as

$$E_3 = \int_{-\infty}^{\infty} dy \int_{0}^{h(y)} dz\left[\sigma_{sl}(y,z) - \sigma_{sg}(y,z)\right], \tag{25}$$

which becomes

$$E_3 = \int_{-\infty}^{\infty} dy \int_{0}^{h(y)} dz \frac{\sigma_{sl}(y,z) - \sigma_{sg}(y,z)}{\sigma_{lg}} \equiv \int_{-\infty}^{\infty} dy \int_{0}^{h(y)} dz \Gamma(y,z), \tag{26}$$

in the units of length and energy chosen previously. In Fourier space the condition $\delta(e + E_3) = 0$ becomes now to $\tilde{\Gamma}(q) - \sqrt{1+q^2}\tilde{h}(q) = 0$, or

$$\tilde{h}(q) = \frac{\tilde{\Gamma}(q)}{\sqrt{1+q^2}}, \tag{27}$$

where

$$\tilde{\Gamma}(q) = \frac{1}{\sqrt{2\pi}} \int_{-\infty}^{\infty} dy e^{iqy} \Gamma(h(y), y). \tag{28}$$

Transforming Eq. (28) into real space, we get Eq. (5). Here we should point out that Eq. (5) has been derived by assuming that gravitational force enables one to find a relation between the contact line profile and the spreading force. In our model, we have not taken into account gravity because momentum conservation equation, where the gravity would naturally manifest itself, has been neglected. However, it is possible to show that confining the liquid between two plates generates a cutoff, which mimics the effect of gravity.

**B. Mesoscopic surface tension**

Using this approach, we can write the free energy per unit area $\Delta A$ relative to the energy of the ground state as

$$\frac{F}{\Delta A} = \int_{-\infty}^{\infty} dy \left[ \tilde{f}(\phi) + \frac{\gamma}{2} \left( \frac{d\phi}{dy} \right)^2 \right] = \int_{-\infty}^{\infty} dy e(y), \tag{29}$$

where $\tilde{f}(\phi) = f(\phi) - f_0$ is the difference between free energy density $f(\phi)$ and the free energy density $f_0$ corresponding to the domain wall (kink) solution to the $\phi^4$ field theory. Namely, in equilibrium, the phase field $\phi$ satisfies the Euler-Lagrange equation

$$\frac{\delta}{\delta \phi(y)} \frac{F}{\Delta A} = \frac{\partial}{\partial \phi(y)} \frac{F}{\Delta A} - \nabla \cdot \frac{\partial}{\partial \nabla \phi(y)} \frac{F}{\Delta A} = -\gamma \frac{d^2 \phi}{dy^2} + \frac{d\tilde{f}}{d\phi}. \tag{30}$$

By multiplying both sides with $d\phi/dy$ and integrating we obtain

$$\frac{d\phi}{dy} = \pm \sqrt{\frac{2\tilde{f}}{\gamma} + C}. \tag{31}$$

The integration constant $C$ is zero because at $y = \pm\infty$, $\tilde{f} = 0$ and $d\phi/dy = 0$. The effective surface tension can be calculated as the excess free energy due to the domain wall

$$\sigma = \int_{-\infty}^{\infty} e(y) dy = 2 \int_{-\infty}^{\infty} \tilde{f}(\phi(y)) dy = 2\sqrt{\frac{\gamma}{2}} \int_{\phi_1}^{\phi_2} \sqrt{\tilde{f}(\phi)} d\phi. \tag{32}$$

Using the last part of Eq. (32) we can express the liquid-gas surface tension as

$$\sigma_{lg} = \frac{4}{3} \sqrt{\frac{\gamma}{2}}, \tag{33}$$

the solid-liquid surface tension

$$\sigma_{sl} = \sqrt{\frac{\gamma\kappa}{2}}(\phi_0 - A)^2 + \frac{1}{3}\sqrt{\frac{\gamma}{2}}(\phi_0^3 - 3\phi_0 + 2) \tag{34}$$

and the solid-gas surface tension

$$\sigma_{sg} = \sqrt{\frac{\gamma\kappa}{2}}(\phi_1 - A)^2 + \frac{1}{3}\sqrt{\frac{\gamma}{2}}(-\phi_1^3 + 3\phi_1 + 2) \tag{35}$$

from the phase field model. In the above equations $\phi_0$ and $\phi_1$ are the equilibrium values of the phase field at the solid-liquid and solid-gas boundaries, correspondingly. We can determine the $\phi_0$ and $\phi_1$ in two different ways. First, we require that the equilibrium domain-wall solutions of the fluid phase $\phi(x) = \tanh(\pm x/\sqrt{2\gamma} + \delta)$ and solid phase $\phi(x) = A + c\exp(\pm\sqrt{\kappa/\gamma}\,x)$ are continuous at the phase boundaries. The other constant can be determined by requiring that the first derivative of the phase field $d\phi/dx$ is continuous at the boundary or by requiring that the free energy is minimized, which corresponds minimizing the surface tension $d\sigma/dc = 0$. Both methods yield the same values for the phase field in the solid-liquid boundary given by

$$\phi_0 = \frac{-\sqrt{2\kappa} + \sqrt{2\kappa + 4 + 4A\sqrt{2\kappa}}}{2} \tag{36}$$

and in the solid-gas boundary given by

$$\phi_1 = \frac{\sqrt{2\kappa} - \sqrt{2\kappa + 4 - 4A\sqrt{2\kappa}}}{2}. \tag{37}$$

Note that these values do not depend on the parameter γ.

**Figure Captions**

**Figure 1.**

(a) A typical meniscus configuration as determined by the condition $\phi(x, y, z) = 0$ for the case with a homogenous wall at *x*=0 and a stripe defect on the wall at *x*=$L_x$+1.
(b) The initial boundary conditions in *xz* plane for the system. The initial phase field variable is $\phi$=1, -1 and *A* for liquid, gas and solid phases, respectively. At the solid and gas boundaries, $\nabla\mu = 0$ and at the liquid boundary, $\nabla\mu = 0$ and $\mu(z=0) = 0 = const.$ for the cases of relaxation of the contact line and capillary-driven wetting, respectively.
(c) The initial boundary conditions in the *yz* plane for the system. Periodic boundary condition is used in the *y* direction. The initial phase field $\phi$=*A* and *B* for the wall and the stripe defect, respectively. At the solid boundary, $\nabla\mu = 0$.

(d) A schematic illustration of a contact line configuration, showing the definitions of the distortion lengths $\xi_z$, $\xi_u$, and $\xi_l$ as measured from the middle of the defect, and the healing length of the distortion $\xi_y$.

**Figure 2.** The height of the contact line *H*(*y*, *t*) for capillary-driven wetting as a function of time for different wetting properties in the case of homogenous walls at $x=0$ and at $x=L_x+1$. These data are for a relatively narrow tube with $L_x \times L_y \times L_z = 10 \times 64 \times 64$, where the crossover to the expected Washburn law $t^{0.5}$ occurs relatively rapidly.

**Figure 3.** The case of capillary-driven wetting for a stripe defect at $x=L_x+1$ with $A=0.5$, $B=-0.5$.
(a) The evolution of the contact line for defect width $L=10$.
(b) Contact line configurations $H(y)$ in the equilibrium state for different defect widths.
(c) The distortion lengths $\xi_z$, $\xi_u$, $\xi_l$, and the healing length $\xi_y$ plotted against the defect width $L$.
(d) A cross-section of the equilibrium meniscus surface in the $xz$ plane as measured in the middle of the defect ($y=0$).

**Figure 4.** $\theta_B$ plotted against defect width for different wetting parameters.

**Figure 5.** The case of capillary-driven wetting for a stripe defect at $x=L_x+1$.
(a), (b) Contact line configurations $H(y)$ in the equilibrium state for different defect widths for $A=0.5$, $B=0$ and $A=0$, $B=-0.5$, respectively.
(c), (d) A cross-section of the equilibrium meniscus surface in the $xz$ plane as measured in the middle of the defect ($y=0$) for $A=0.5$, $B=0$ and $A=0$, $B=-0.5$, respectively. The corresponding contact angles $\theta_B$ as shown in Figure 4.

**Figure 6.** The case of capillary-driven wetting for two stripe defects ($L=20$) at $x=L_x+1$ with $A=0$, $B=-0.5$. The evolution of the contact line $H(y, t)$ for different distances $d$ between defects: (a) $d=4$, (b) $d=8$, (c) $d=12$ and (d) $d=20$.

**Figure 7.** The case of relaxation of the contact line to its equilibrium configuration for a stripe defect at $x=L_x+1$ with $A=0$, $B=-0.5$. (a) The evolution of the contact line $H(y, t)$ for defect width $L=20$. (b) Contact line configurations $H(y)$ in the equilibrium state for different defect widths. (c) A cross-section of the equilibrium meniscus surface in the $xz$ plane as measured in the middle of the defect at $y=0$.

**Figure 8.** Analytical results of contact line configurations in equilibrium state with defect widths of the case $A=0$, $B=-0.5$ for different methods: (a) Method 1 [see Eq. (7)], (b) Method 3 [see Eq. (10)]. (c) Comparison of contact line configurations for Methods 1 and 3.

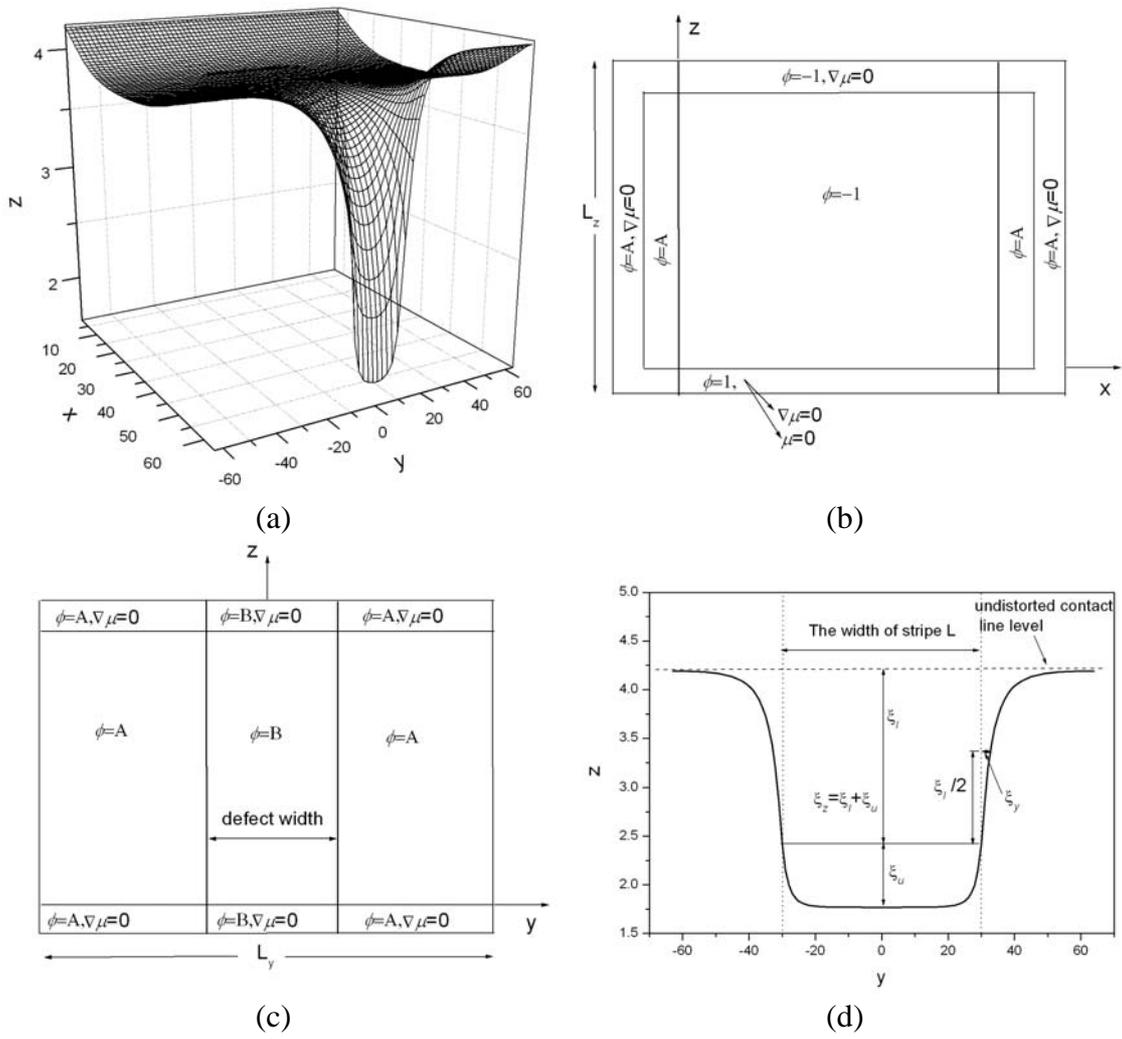

(a)          (b)

(c)          (d)

**Figure 1.**

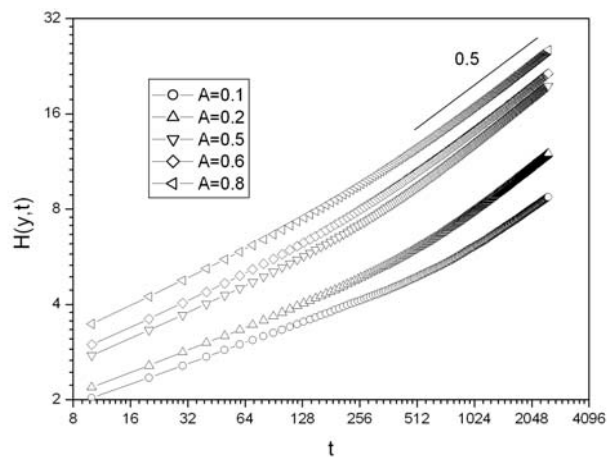

**Figure 2.**

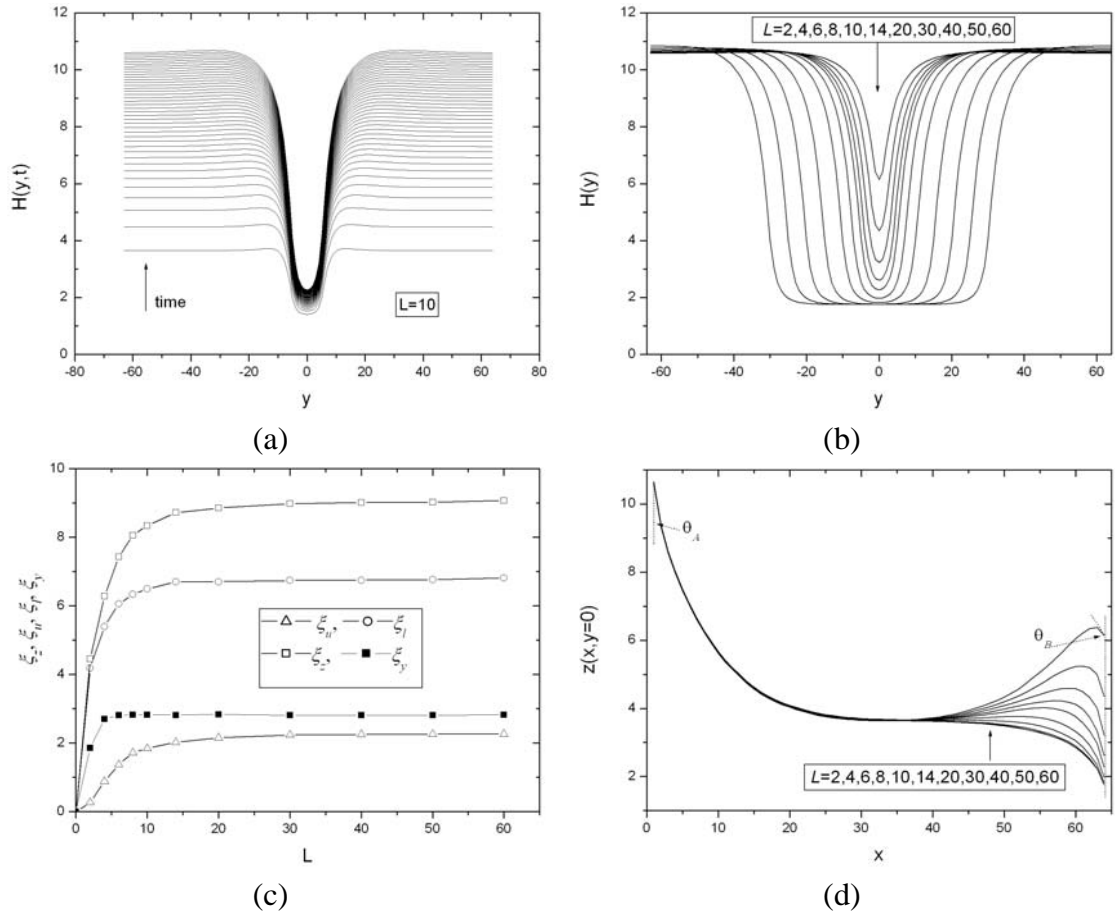

**Figure 3.**

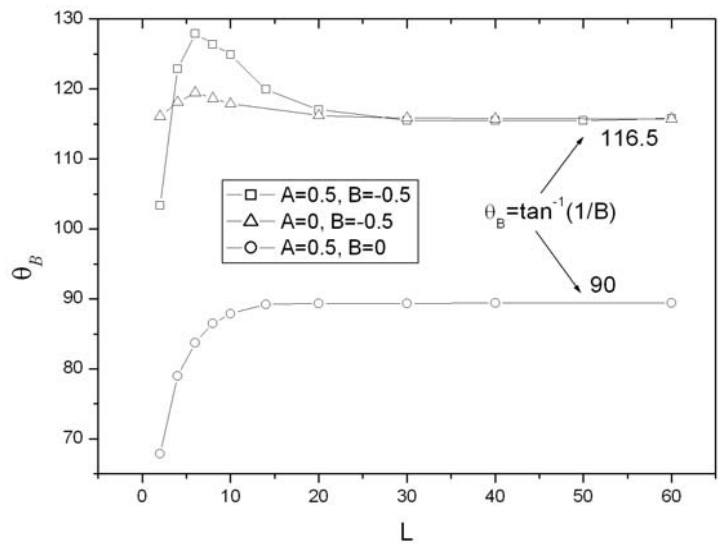

**Figure 4.**

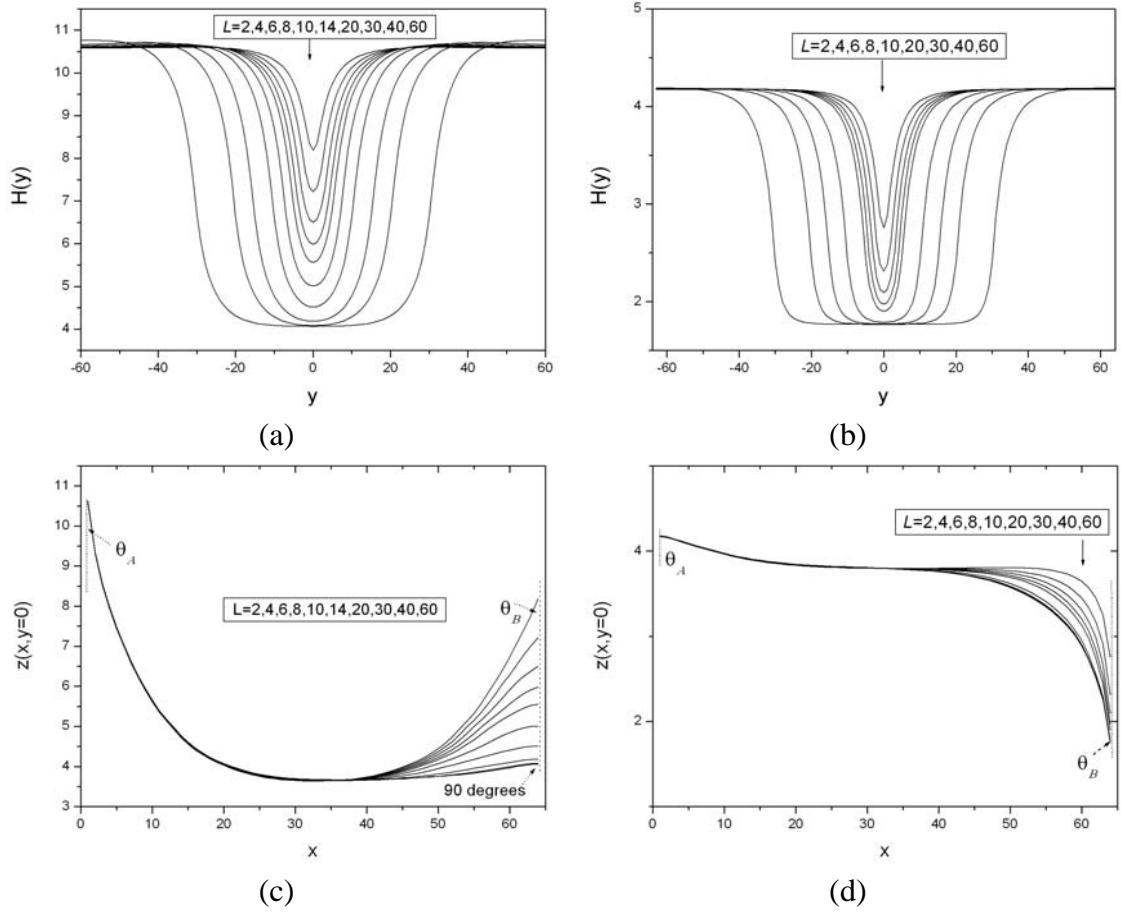

**Figure 5.**

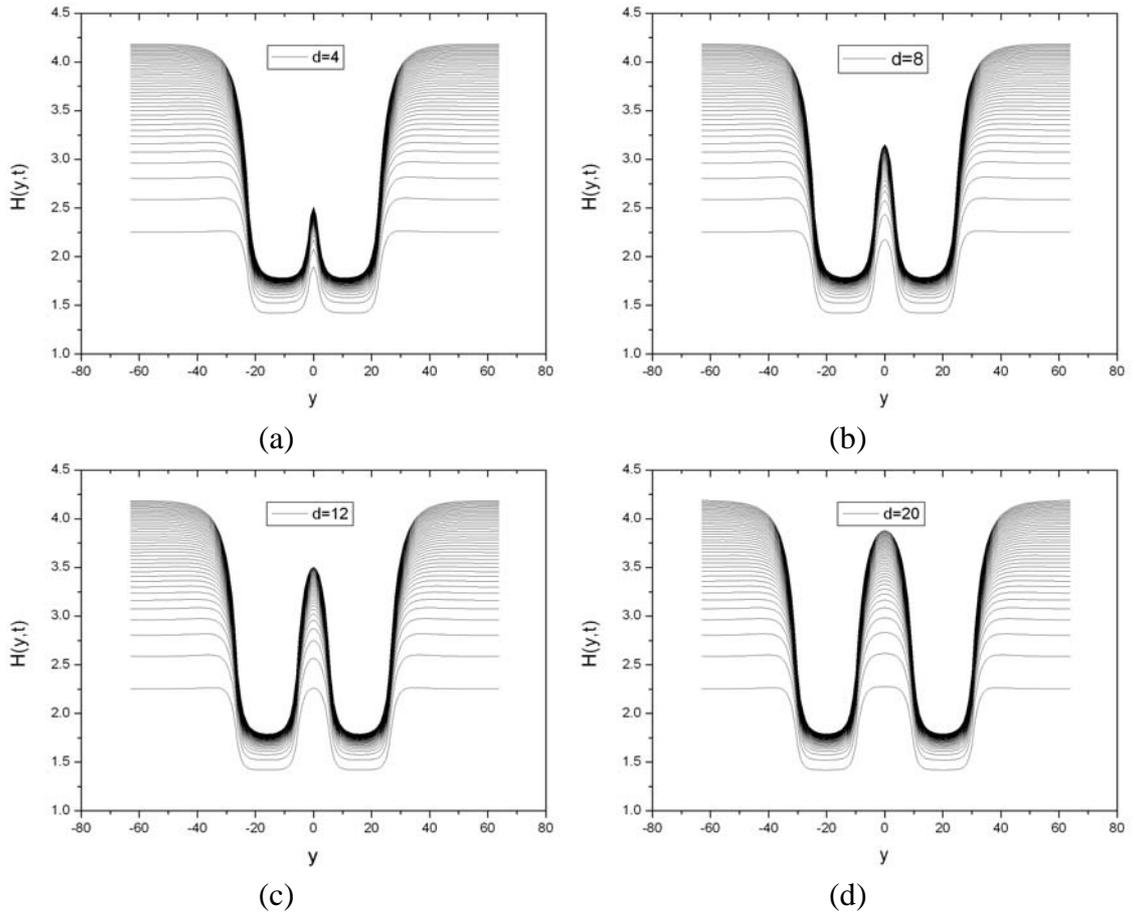

**Figure 6.**

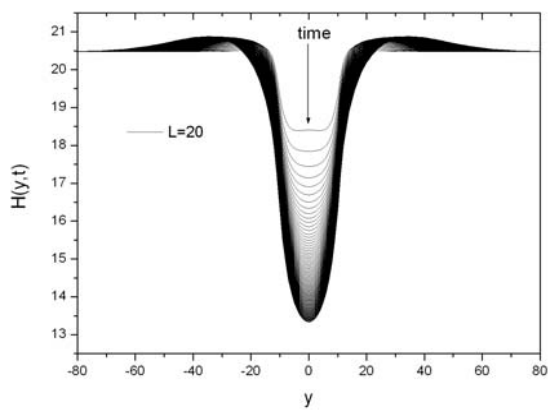

(a)

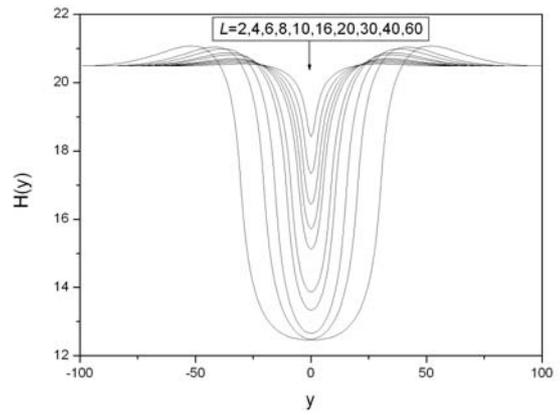

(b)

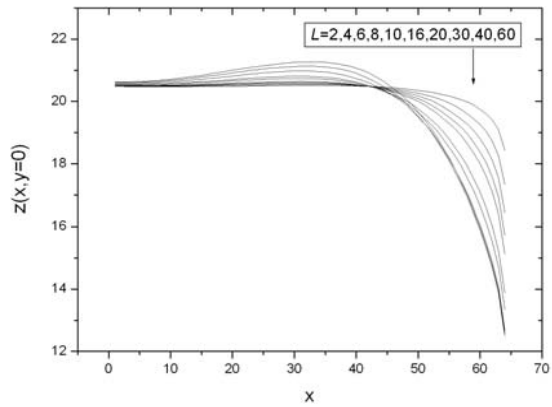

(c)

**Figure 7.**

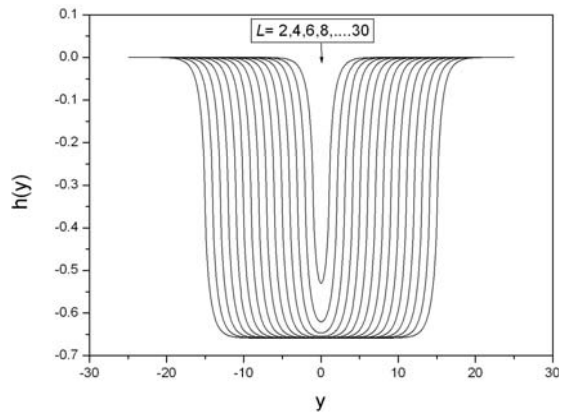

(a)

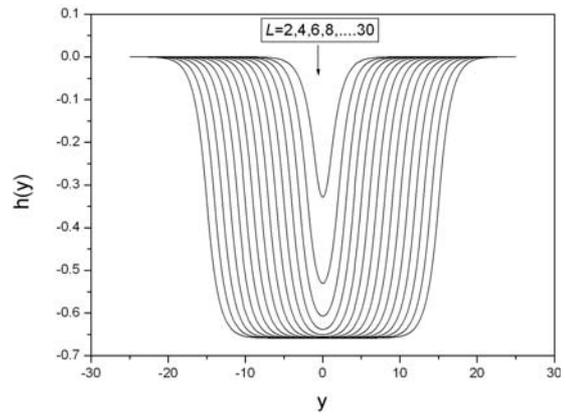

(b)

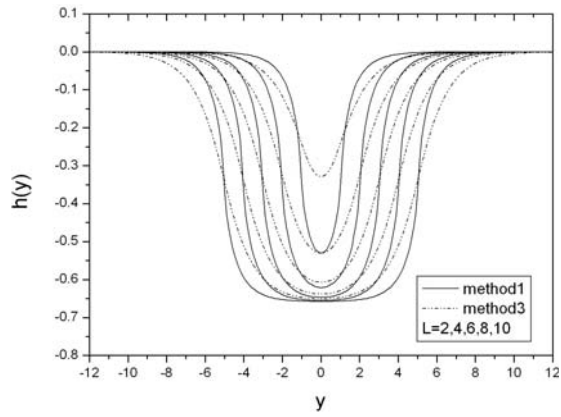

(c)

**Figure 8.**